\documentclass[12pt]{article}
\usepackage{epsfig}
\begin{document}

\begin{center}
{\large\bf SENSITIVITY OF REACTION CROSS SECTIONS TO HALO NUCLEUS DENSITY DISTRIBUTIONS} \\
\bf{G.~D.~Alkhazov,~V.~V.~Sarantsev$\hspace*{2pt}^*$}\\
{\it Petersburg Nuclear Physics Institute of National Research
Centre\\"Kurchatov Institute", Gatchina, Russia}
\end{center}

\vspace{7mm} \noindent In order to clear up the sensitivity of
the nucleus--nucleus reaction cross sections $\sigma_R$ to the
nuclear matter distributions in exotic halo nuclei, we have
calculated the values of $\sigma_R$ for scattering of $^6$He,
$^{11}$Li, and  $^{19}$C nuclei on several nuclear targets at the
energy of 0.8 GeV/nucleon. The calculations were performed in the
"rigid target" approximation to the Glauber theory, different
shapes of the nuclear density distributions in $^6$He, $^{11}$Li,
and $^{19}$C being assumed.
\par
\vspace{0.2cm}
\noindent
{\bf Comments:} 13 pages, 6 figures. Submitted to Physics of Atomic Nuclei\\
\noindent
{\bf Subjects:} Nuclear Theory(nucl-th); High-Energy
Physics-Theory(hep-th)
\vspace{1cm} \noindent

\noindent
{\bf 1. INTRODUCTION}\\
Reaction cross sections $\sigma_R$ serve as one of the main
sources of information on sizes of exotic halo nuclei. The root
mean square (rms) radius of an exotic nucleus is determined by
comparing the experimental reaction cross section for
nucleus-nucleus scattering involving the nucleus of interest with
theoretical model predictions. At the same time, the reaction
cross sections depend not only on the rms nuclear matter radius,
but also to some extent on the radial shape of the studied
nucleus. Bush with coauthors [1] investigated the sensitivity of
reaction cross sections to $^{11}$Li density distributions, and
they concluded that $^{11}$Li-target cross sections at fixed rms
matter radii retain a significant sensitivity to higher radial
moments of the assumed $^{11}$Li density which is quite different
for light and heavy targets. According to their consideration,
reaction cross sections for nucleus--nucleus scattering are
significantly more sensitive to the matter density at the nuclear
periphery than those for proton--nucleus scattering. Therefore,
measuring reaction cross sections for nucleus-nucleus and
nucleus-proton scattering one can determine the rms nuclear matter
radius and obtain information on the radial shape of the studied
nucleus. However, calculations of cross sections in [1] were
performed using the Glauber theory in the optical limit
approximation, which as is well known [2] significantly
overestimates nucleus-nucleus reaction cross sections, especially
for the case of halo nuclei. For this reason, it is not clear
whether the conclusion of Bush {\it et al.} [1] that
nucleus-nucleus cross sections are more sensitive to the nuclear
periphery than nucleus-proton cross sections reflects the real
physical case, or this conclusion was made due to the optical
approximation to the Glauber theory used in their calculations.
Bush {\it et al.} pointed out that it would be important to repeat
similar calculations using a more accurate approach for
calculations of the reaction cross sections.

In the present work, we investigate the sensitivity of reaction
cross sections to the shape of the nuclear matter distribution by
performing calculations of reaction cross sections for scattering
of halo nuclei $^6$He, $^{11}$Li and $^{19}$C on different nuclear
targets at the energy of 0.8 GeV/nucleon. The calculations of
reaction cross sections were performed using the Glauber theory
within the ``rigid target'' approximation [3], that is, at first
the amplitude of scattering of one nucleon on the nuclear target
was calculated, and then this amplitude was used in the
calculations of the cross sections for scattering of an exotic
nucleus consisting of several nucleons. As was shown in [2], the
reaction cross sections for scattering of exotic nuclei on nuclear
targets calculated within the rigid target approximation are very
close to those calculated with the exact Glauber formula.

\noindent
{\bf 2. CROSS SECTION CALCULATIONS}\\
The main formulas used in the present paper for calculations of
the reaction cross sections $\sigma_R$ for nucleus--nucleus
scattering are given below:
\begin{equation}
\sigma_R = \sigma_{tot} - \sigma_{el},
\end{equation}
\begin{equation}
\sigma_{tot} = \frac{4\pi}{k} {\rm Im} [F_{el}(0)], \hspace{1cm}
\sigma_{el} = \frac{2\pi}{k^2} \int \vert F_{el}({\bf q}) \vert^2
q dq,
\end{equation}
\begin{eqnarray}
F_{el}({\bf q}) & = & \frac{ik}{2\pi} \int d^2b ~ {\rm exp} ~
(i{\bf
qb}) ~ d^3r_1 d^3r_2...d^3r_A \cdot \rho({\bf r}_1,...,{\bf r}_A) \quad \times \\
\nonumber & \times & \left\{1 - \prod_{j=1}^A \left[1 -
\Gamma({\bf b} - {\bf s}_j) \right] \right\} ,
\end{eqnarray}
\begin{equation}
\Gamma({\bf b}) = \int d^3r_1 d^3r_2...d^3r_{A_{\rm T}} ~
\rho_{\rm T}({\bf r}_1,...,{\bf r}_{A_{\rm T}}) \times \left\{1 -
\prod_{m=1}^{A_{\rm T}} \left[1 - \gamma({\bf b} - {\bf s}_m)
\right] \right\} .
\end{equation}

Here, $\sigma_{tot}$ is the total cross section for
nucleus-nucleus scattering, $\sigma_{el}$ is the integrated cross
section for elastic nucleus-nucleus scattering, $k$ is the value
of the wave vector of the incident exotic nucleus, $F_{el}({\bf
q})$ is the amplitude of elastic nucleus--nucleus scattering, {\bf
q} is the momentum transfer, ${\bf b}$ is the impact parameter
vector, $\rho({\bf r}_1,...,{\bf r}_A)$ and $\rho_{\rm T}({\bf
r}_1,...,{\bf r}_{A_{\rm T}})$ are the many-body densities
correspondingly of the studied exotic nucleus and the target
nucleus, ${\bf r}_1,...,{\bf r}_A$, ${\bf r}_1,...,{\bf r}_{A_{\rm
T}}$ and ${\bf s}_1,...,{\bf s}_A$, ${\bf s}_1,...,{\bf s}_{A_{\rm
T}}$ stand for the radius vectors of the nucleons in these nuclei
and their transverse coordinates, $A$ and $A_{\rm T}$ are the
total numbers of nucleons in the exotic and target nuclei,
$\Gamma({\bf b})$ is the profile function of scattering of one
nucleon of the studied exotic nucleus on the target nucleus, and
$\gamma({\bf b})$ is the profile function of the nucleon--nucleon
interaction. Spin-independent isospin-averaged amplitude of the
free nucleon--nucleon ($NN$) scattering was employed, the
traditional high-energy parametrization of this amplitude and the
corresponding profile function
\begin{equation}
\gamma({\bf b}) = \frac{\sigma_{NN}(1 - i\epsilon_{NN})}{2}  ~
\frac{1}{2 \pi \beta_{NN}} ~ {\rm exp}\Big(- \frac{{\bf b}^2}{2
\beta_{NN}}\Big)
\end{equation}
being taken with the same parameters as in [1]: the total cross
section $\sigma_{NN} =$ 41 mb and the amplitude slope $\beta_{NN}
=$ 0 fm$^2$. The ratio of the real to imaginary part
$\epsilon_{NN}$ of the $NN$ scattering amplitude has very little
influence on the calculated value of $\sigma_R$. Within the
optical limit Glauber model, this parameter has no influence on
$\sigma_R$ at all. In our calculations we set $\epsilon_{NN} = 0$.

\noindent
{\bf 3. DENSITY DISTRIBUTIONS}\\
In the present calculations it was assumed that the $^6$He,
$^{11}$Li, and $^{19}$C nuclei consist of a nuclear core of four,
nine, and 18 nucleons, correspondingly, the number of halo
neutrons being correspondingly two, two, and one. The many-body
densities in the projectile exotic nucleus and in the target
nucleus were presented as products of one-body densities:
\begin{equation}
\rho({\bf r}_1,...,{\bf r}_A) = \prod_{j=1}^{A_{\rm c}} \rho_{\rm
c} ({\bf r}_j) \times \prod_{j=A_{\rm c} + 1}^{A_{\rm c} + A_{\rm
h}} \rho_{\rm h} ({\bf r}_j),
\end{equation}
\begin{equation}
\rho_{\rm T}({\bf r}_1,...,{\bf r}_{A_{\rm T}}) =
\prod_{j=1}^{A_{\rm T}} \rho_{\rm T} ({\bf r}_j).
\end{equation}

Here, $\rho_{\rm c} ({\bf r}_j)$ and $\rho_{\rm h} ({\bf r}_j)$
are the one-body densities of the core and halo of the exotic
nucleus, correspondingly; $\rho_{\rm T} ({\bf r}_j)$ is the
one-body density of the target nucleus; $A_{\rm c}$ and $A_{\rm
h}$ are the numbers of the core and halo nucleons ($A_{\rm c} +
A_{\rm h} = A$).

As in [1], the matter density distributions in the core were
described by Gaussian distributions
\begin{equation}
\rho_{\rm c} ({\bf r}_j) = (3/2\pi {R_{\rm c}}^2)^{3/2} ~ {\rm
exp} (- 3 {{\bf r}_j}^2 /2{R_{\rm c}}^2),
\end{equation}
whereas the density distributions in the halo were described by a
1$p$-shell harmonic oscillator-type function
\begin{equation}
\rho_{\rm h} ({\bf r}_j) = (5/3)(5/2\pi R^2_{\rm h})^{3/2} ~ ({\bf
r}_j/R_{\rm h})^2 ~ {\rm exp} (-5{{\bf r}_j}^2 /2{R_{\rm h}}^2).
\end{equation}
Here, $R_{\rm c}$ and $R_{\rm h}$ are the root-mean-square (rms)
radii of the core and halo matter density distributions.
Calculations with the halo density distributions containing long
density ``tails'' (to be discussed later) were also performed.
Note that the rms radius $R_{\rm m}$ of the total matter density
distribution is connected with the core and halo radii $R_{\rm c}$
and $R_{\rm h}$ as
\begin{equation}
R_{\rm m} = \left[(A_{\rm c}R_{\rm c}^2 + A_{\rm h}R_{\rm h}^2)/A
\right]^{1/2}.
\end{equation}

We used the matter density distributions in the target nuclei
$^4$He, $^{12}$C, $^{28}$Si, $^{58}$Ni, and $^{208}$Pb the same as
those in [4].

As is known, no nucleon correlations are taken into account in
cross-section calculations in the optical limit. When performing
calculations of the cross sections in the rigid target
approximation, we did not take the nucleon correlations into
account either.

\noindent
{\bf 4. RESULTS AND DISCUSSION}\\
At first, using the Glauber formulas in the optical limit
approximation we repeated the calculations of Bush {\it et al.}
[1] of reaction cross sections $\sigma_R$ for scattering of
$^{11}$Li on protons and nuclear targets $^{12}$C and $^{208}$Pb
at several fixed $^{11}$Li rms matter radii $R_{\rm m}$ with
different ratios $R_{\rm h}$/$R_{\rm c}$. The calculated cross
sections are shown in Fig. 1 versus the ratio $\epsilon = < r^4 >
/ < r^2 >^2$. The results of our calculations are practically the
same as in [1]. (A small difference between the cross sections
calculated in our study and in [1] is evidently due to slightly
different target nuclear matter distributions used in the
calculations.) In addition to the nuclear targets $^{12}$C and
$^{208}$Pb, considered by Bush {\it et al.}, we also performed
calculations for the $^4$He target.  It is seen that the
dependence of $\sigma_R$ upon $\epsilon$ is essentially different
for light and heavy targets. According to [1], these results
indicate that reaction cross sections for heavy targets have
higher sensitivity to valence nucleons of exotic nuclei than that
for light targets.
\begin{figure}[ht]
\centerline{\epsfig{file=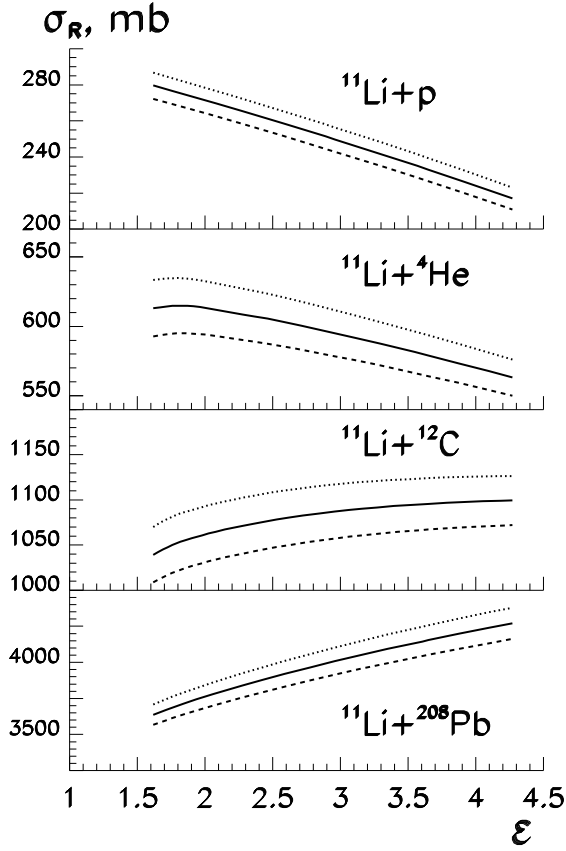,width=0.60\textwidth}}
\caption{Variation of reaction cross sections $\sigma_R$ with
$\epsilon = < r^4 > / < r^2 >^2$ at fixed total rms matter radii
$R_{\rm m}$ of $^{11}$Li equal to 2.9 fm (dashed lines), 3.0 fm
(solid lines), and 3.1 fm (dotted lines). The cross sections are
calculated in the optical limit approximation to the Glauber theory
for scattering of $^{11}$Li on hydrogen, $^4$He, $^{12}$C, and
$^{208}$Pb targets.}
\end{figure}

Then, we repeated similar calculations using the Glauber theory
within the rigid target approximation (Fig. 2). The cross sections
calculated in the rigid target approximation are somewhat
different from those calculated in the optical limit
approximation. At the same time, for the given $R_{\rm m}$ values,
the variations of the cross sections as a function of $\epsilon$
are more or less similar in both cases. With $\epsilon$
increasing, the reaction cross sections $\sigma_R$ for light
targets (protons and $^4$He) decrease, while for heavy targets
($^{208}$Pb) they increase. For $^{11}$Li-$^{12}$C scattering, the
cross section calculated in the optical limit approximation
increases by a few percent with $\epsilon$ increasing, while the
cross section calculated in the rigid target approximation is
almost constant, the variation of $\sigma_R$ with $\epsilon$ being
very small.

\begin{figure}[h]
\centerline{\epsfig{file=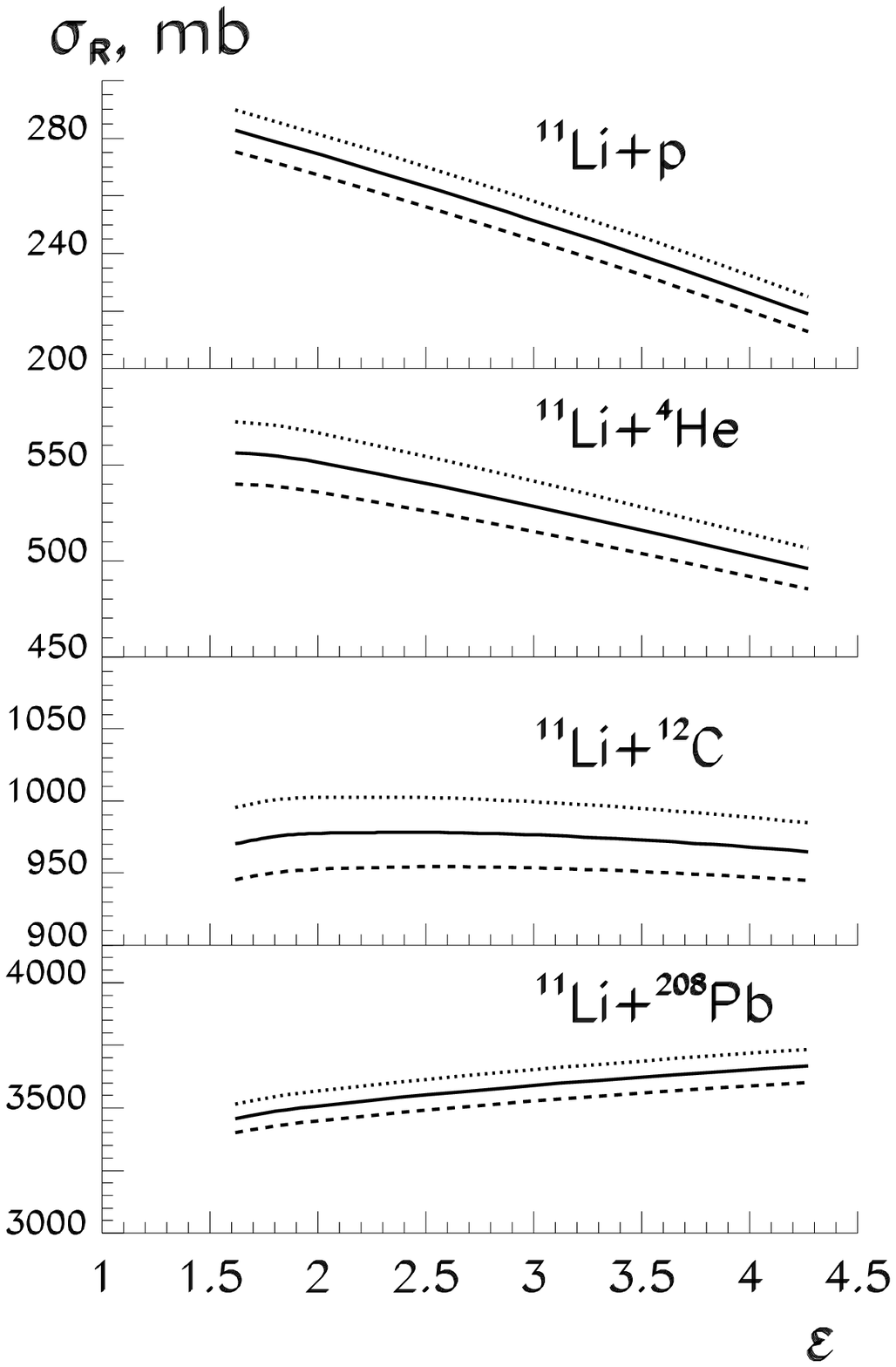,width=0.60\textwidth}}
\caption{Variation of reaction cross sections $\sigma_R$ with
$\epsilon$. The same as in Fig. 1 for the cross sections calculated
in the rigid target approximation.}
\end{figure}

We also carried out calculations of the reaction cross sections
$\sigma_R$ for scattering of halo nuclei $^6$He and $^{19}$C on
protons and nuclear targets $^4$He, $^{12}$C and $^{28}$Si. As
compared to $^{11}$Li, the considered nuclei $^6$He and $^{19}$C
have different relative amount of halo neutrons. The results of
calculations (Figs. 3 and 4) show that here also with $\epsilon$
increasing the cross sections $\sigma_R$ decrease  in the case of
light target nuclei (protons and $^4$He), while they increase for a
heavier nucleus $^{28}$Si.
\begin{figure}[h]
\centerline{\epsfig{file=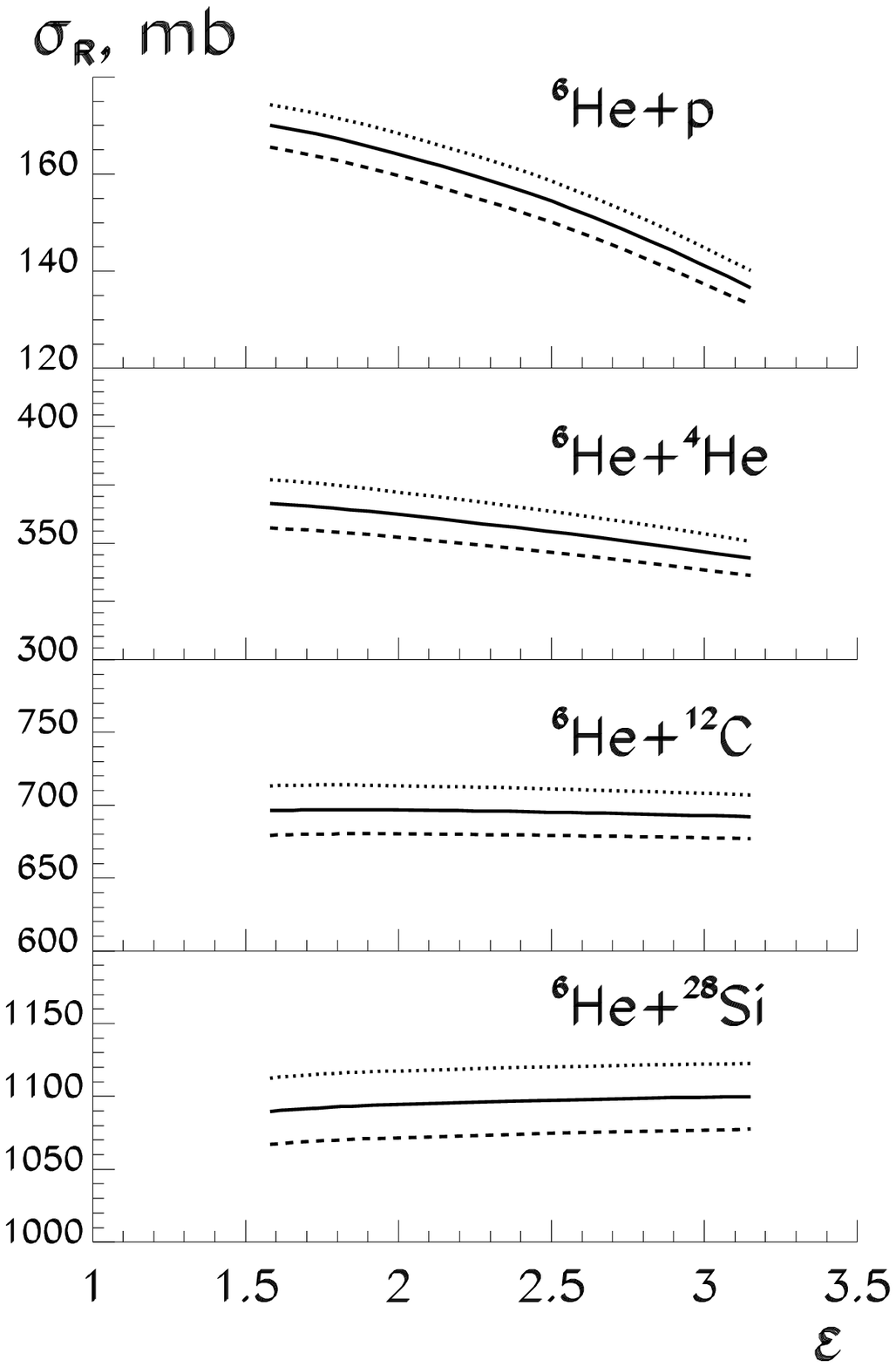,width=0.60\textwidth}}
\caption{Variation of reaction cross sections $\sigma_R$ with
$\epsilon$ at fixed total rms matter radii $R_{\rm m}$ of $^6$He
equal to 2.4 fm (dashed lines), 2.5 fm (solid lines), and 2.6 fm
(dotted lines). The cross sections are calculated in the rigid
target approximation for scattering of $^6$He on hydrogen, $^4$He,
$^{12}$C, and $^{28}$Si targets.}
\end{figure}
The performed calculations confirm the conclusions of Bush {\it et
al.} that the cross sections $\sigma_R$ at fixed rms radii of halo
nuclei retain a significant sensitivity to higher radial moments of
the nuclear density which is different for light and heavy targets.
This means that information about the rms radius can only be
meaningfully extracted from the measured cross section if some form
is assumed for the radial density distribution of the studied
nucleus. At the same time, analyzing reaction cross sections
measured for several nuclear targets of different size one can more
precisely asses the rms nuclear matter radius and also get
information on the radial shape of the studied nucleus. (Note that
in the case of heavy targets Coulomb dissociation of the exotic
nuclei should be also taken into account.)
\begin{figure}[h]
\centerline{\epsfig{file=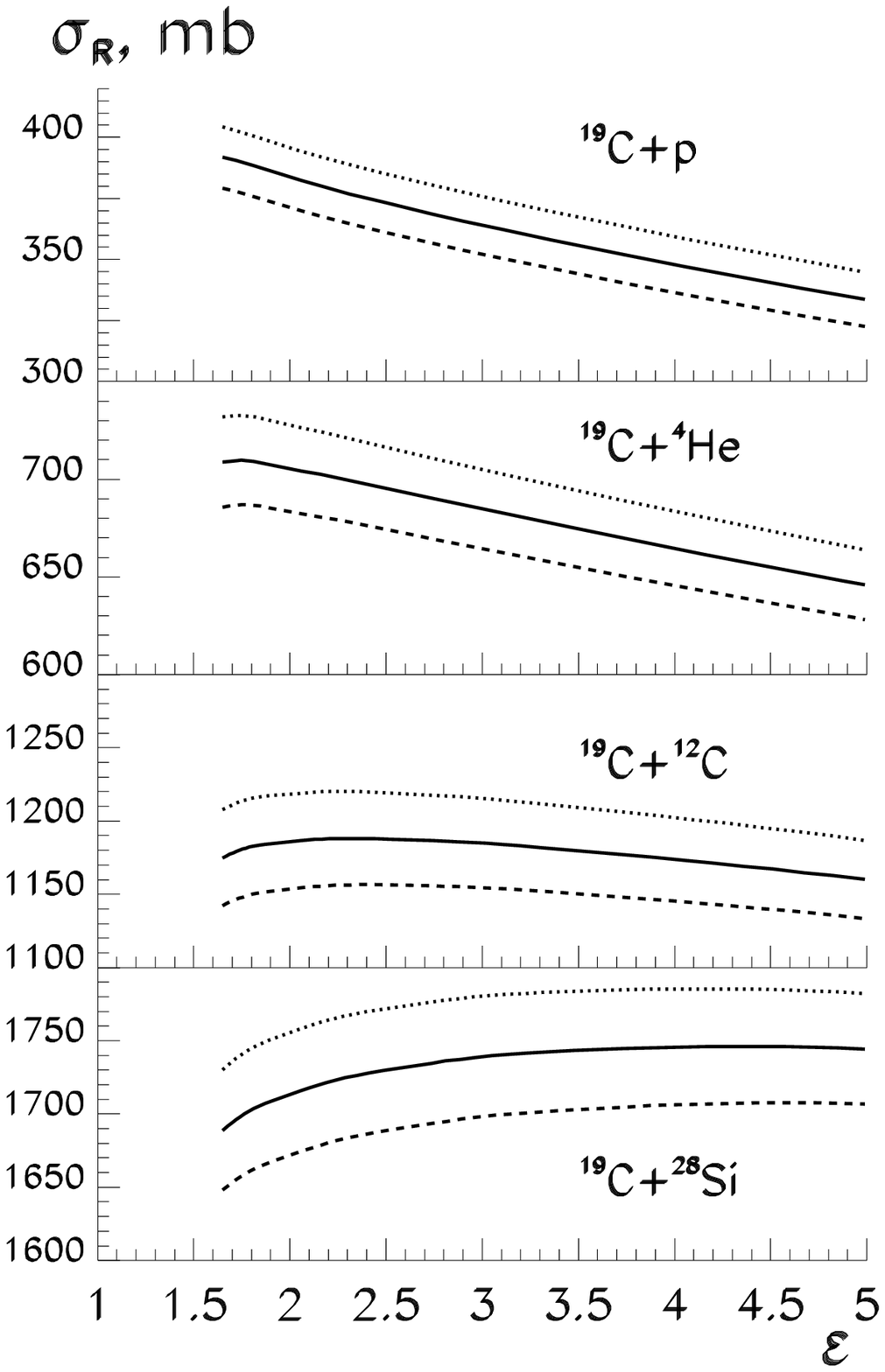,width=0.60\textwidth}}
\caption{Variation of reaction cross sections $\sigma_R$ with
$\epsilon$ at fixed total rms matter radii $R_{\rm m}$ of $^{19}$C
equal to 3.0 fm (dashed lines), 3.1 fm (solid lines), and 3.2 fm
(dotted lines). The cross sections are calculated in the rigid
target approximation for scattering of $^{19}$C on hydrogen, $^4$He,
$^{12}$C, and $^{28}$Si targets.}
\end{figure}

Equation (9) with $R_{\rm h} > R_{\rm c}$ can describe an extended
nucleon distribution of valence nucleons. However, this distribution
at large distances from the nuclear center decreases with the radius
$r$ increasing faster than it is predicted by theory. According to
[5, 6], in addition to the main halo component which can be
described to a first approximation by (9), the halo in exotic nuclei
with low binding energy contains also a long density ``tail'', which
decreases exponentially with the radius $r$ increasing (Figs. 5, 6).
Though such a density tail contains a small amount of matter (of the
order of 1 \%), it can produce a noticeable effect on the total rms
nuclear matter radius. Due to smallness of the density tails, it can
be expected that the sensitivity of the reaction cross sections
$\sigma_R$ to the density tails is relatively poor. In the present
study we investigate the sensitivity of reaction cross sections to
the density tails by performing calculations of $\sigma_R$ using
model nuclear density distributions with a tail and without it.
\begin{figure}[ht]
\centering\epsfig{file=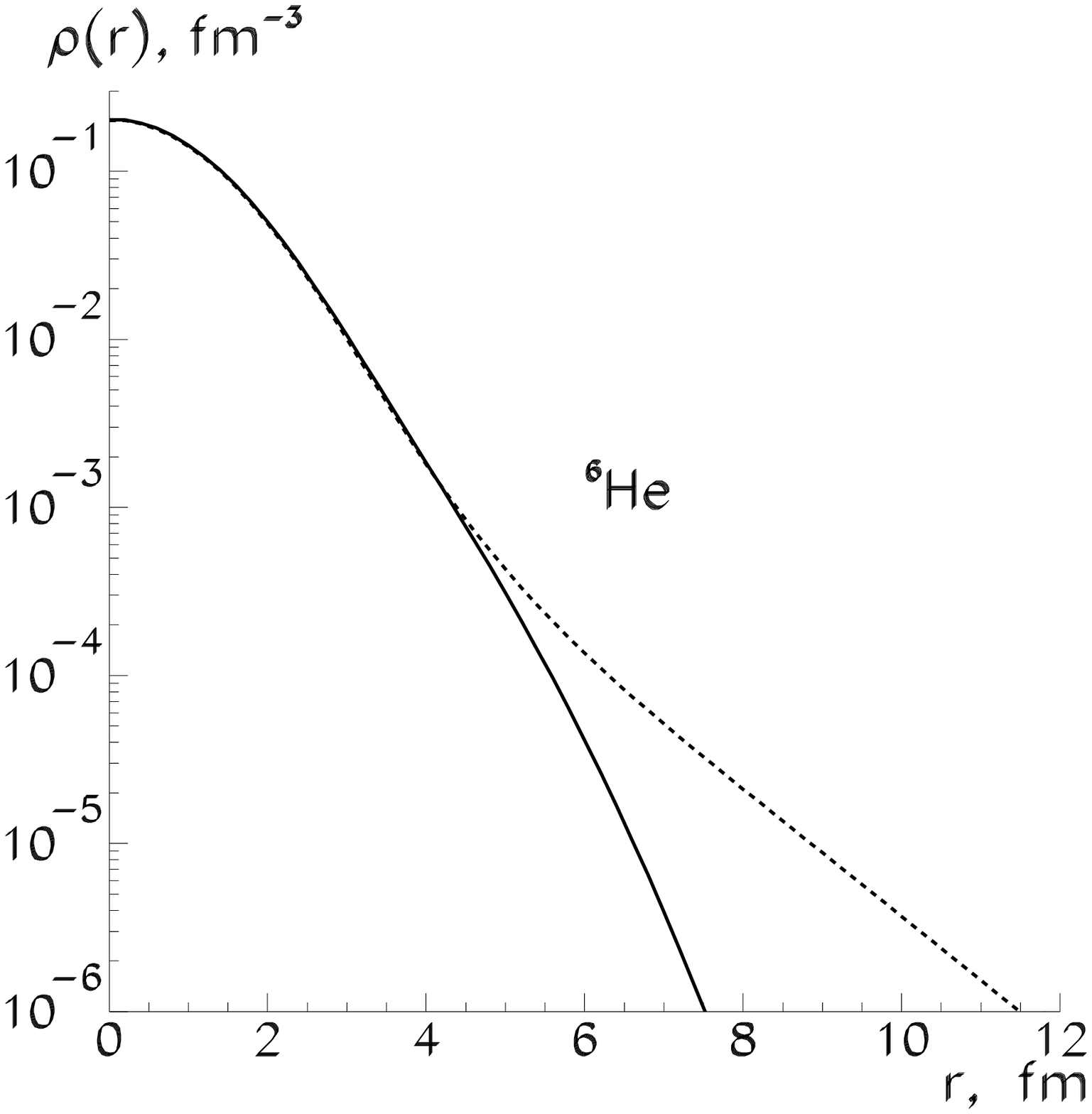,height=80mm} \caption{The
nuclear matter density distribution in $^6$He (applied in the
cross-section calculations) without a density tail (solid curve,
$R_{\rm c}$ = 1.95 fm, $R_{\rm h}$ = 2.88 fm, $R_{\rm m}$ = 2.30 fm)
and with a density tail (dashed curve, see the text).}
\end{figure}

\begin{figure}[ht]
\centering\epsfig{file=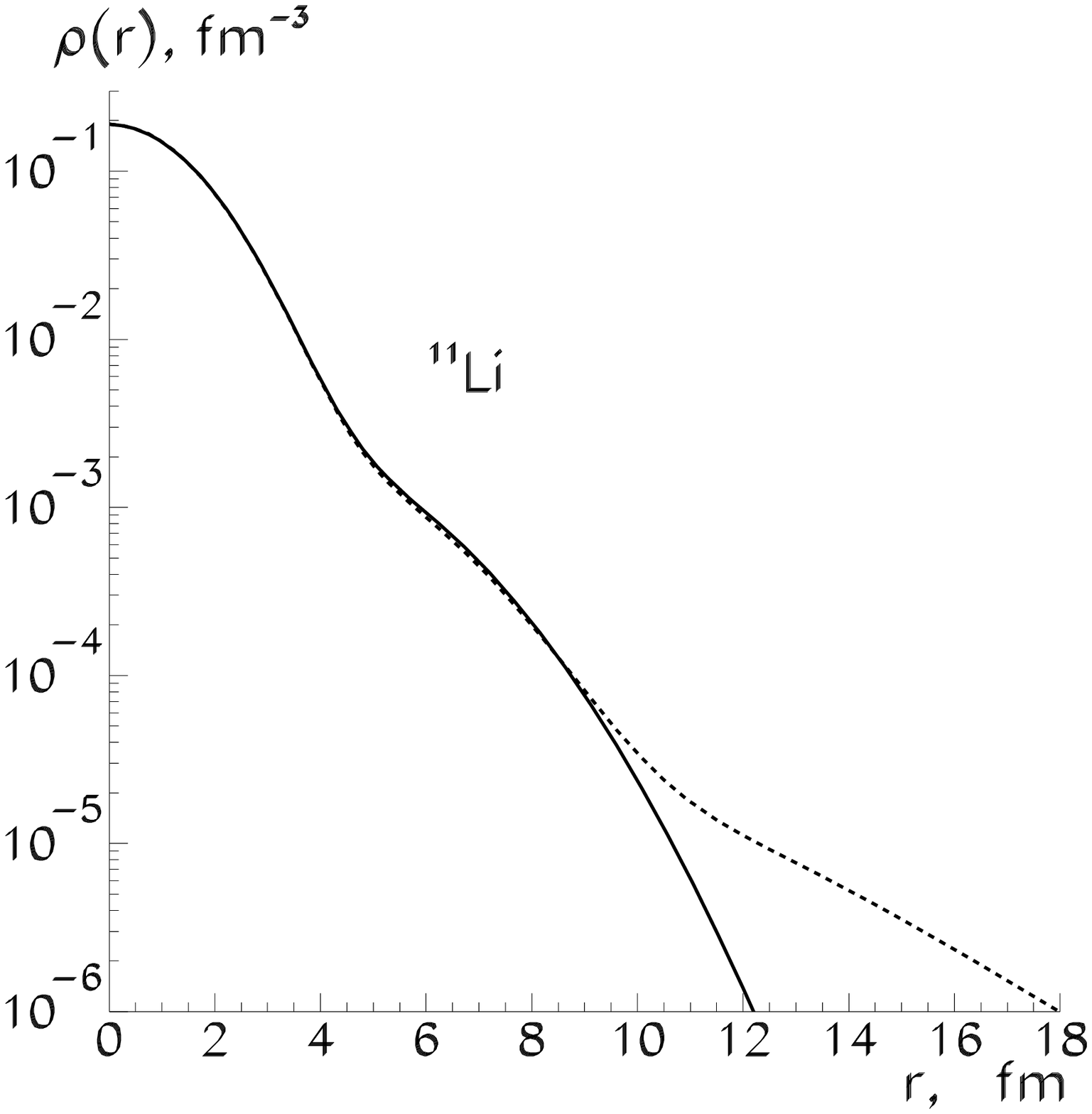,height=80mm} \caption{The
nuclear matter density distribution in $^{11}$Li (applied in the
cross-section calculations) without a density tail (solid curve,
$R_{\rm c}$ = 2.30 fm, $R_{\rm h}$ = 5.86 fm, $R_{\rm m}$ = 3.37 fm)
and with a density tail (dashed curve, see the text).}
\end{figure}

We performed calculations of $\sigma_R$ for scattering of $^6$He and
$^{11}$Li on protons and several nuclear targets. The nuclear
density distributions in $^6$He and $^{11}$Li used in the
calculations are shown in Figs. 5 and 6. The solid curves in these
figures show the density distributions without a tail with the
following parameters: $R_{\rm c}$ = 1.95 fm, $R_{\rm h}$ = 2.88 fm,
$R_{\rm m}$ = 2.30 fm for $^6$He, and $R_{\rm c}$ = 2.30 fm, $R_{\rm
h}$ = 5.86 fm, $R_{\rm m}$ = 3.37 fm for $^{11}$Li (see [7, 8]). The
dashed curves show the same distributions including the tails
corresponding to the theoretical density distributions FC of [5] and
P2 of [6]. Inclusion of these tails increases the rms matter radii
by 0.20 fm and 0.26 fm, correspondingly in $^6$He and $^{11}$Li.

Table 1 presents the relative increase $\Delta \sigma_R$ /
$\sigma_R$ of the calculated reaction cross sections when the
density tails are taken into account. For comparison, we also show
the relative increase $\Delta \sigma'_R$ / $\sigma_R$ of the cross
sections calculated without density tails, but with the increased
halo radii $R_{\rm h}$ so that the total rms matter radii $R_{\rm
m}$ are larger by 0.20 fm and 0.26 fm, correspondingly in $^6$He and
$^{11}$Li.

It is seen that the sensitivity of the reaction cross
sections $\sigma_R$ for nucleus-proton scattering to the density
tails is rather poor, especially when the tail is long but contains
small amount of nuclear matter, as in the case of $^{11}$Li. At the
same time, the sensitivity of $\sigma_R$ to the density tails is
more sizable in the case of nucleus-nucleus scattering for
middle-weight nuclear targets ($^{28}$Si, $^{58}$Ni). In all the
cases, the values of $\Delta \sigma_R$ / $\sigma_R$ are smaller than
those of $\Delta \sigma'_R$ / $\sigma_R$. This means that taking the
density tails into account in the analyses of the reaction cross
sections $\sigma_R$ is important in order to deduce accurate values
of the rms matter radii of the studied nuclei.

\newpage
\vspace{15pt} {\bf Table 1.} Relative increases of the
calculated cross sections ($\Delta \sigma_R$ / $\sigma_R$) when the
density tails (shown in Figs. 5 and 6) are taken into account  and
($\Delta \sigma'_R$ / $\sigma_R$) when the halo radii are increased
(see the text).
\begin{center}
\vskip 5pt
\begin{tabular}{c|c|c} \hline
Interacting & $\Delta \sigma_R$ / $\sigma_R$ & $\Delta \sigma'_R$ / $\sigma_R$ \\ nuclei & \\
\hline
$^6$He + $p$ & 1.8 \% & 3.9 \% \\
$^6$He + $^4$He & 2.9 \% & 4.8 \% \\
$^6$He + $^{12}$C & 3.6 \% & 4.6 \% \\
$^6$He + $^{28}$Si & 3.6 \% & 4.2 \% \\
$^6$He + $^{58}$Ni & 3.5 \% & 4.0 \% \\
$^{11}$Li + $p$ & 0.4 \% & 1.4 \% \\
$^{11}$Li + $^4$He & 1.1 \% & 3.1 \% \\
$^{11}$Li + $^{12}$C & 2.2 \% & 5.2 \% \\
$^{11}$Li + $^{28}$Si & 3.3 \% & 6.1 \% \\
$^{11}$Li + $^{58}$Ni & 4.1 \% & 6.5 \% \\
\hline
\end{tabular}
\end{center}
\vspace{15pt}

\noindent
{\bf 5. CONCLUSION}\\
We have performed calculations of the reaction cross sections
$\sigma_R$ for nucleus-nucleus and nucleus-proton scattering
assuming different model density distributions of the considered
halo nuclei. The calculations were performed within the rigid
target approximation to the Glauber theory, which is more accurate
than the optical limit approximation used in previous calculations
by Bush {\it et al.} [1]. Though the results of our calculations
differ quantitatively from [1], they are in qualitative agreement
with the conclusion of Bush {\it et al.} that the reaction cross
sections for nucleus-nucleus scattering are more sensitive to the
matter density at the nuclear periphery than those for
nucleus-proton scattering. For all the considered cases, the
calculated reaction cross sections $\sigma_R$ for nucleus-nucleus
scattering depend not only on the rms matter radius of the studied
exotic nucleus but also on the shape of its matter distribution.
However, for the case of the $^{12}$C target this dependence is
rather weak, the value of $\sigma_R$ depending basically on the
rms nuclear matter radius. Analyzing the reaction cross sections
for nucleus-proton and nucleus-nucleus scattering for light and
middle-weight target nuclei it is possible to determine the rms
matter radius of the exotic nucleus and to obtain information on
the shape of its matter distribution. Our considerations also have
shown that to deduce accurately the nuclear matter radius from the
measured reaction cross sections it is important to take into
account long matter density tails of the exotic nuclei at the
nuclear far periphery.

\newpage
\noindent
{\bf REFERENCES}\\
1. M. P. Bush, J. S. Al-Khalili, J. A. Tostevin, and R. C.
Johnson, Phys. Rev. C {\bf 53}, 3009 (1996).\\
2. G. D. Alkhazov and A. A. Lobodenko, Yad. Fiz. {\bf 70}, 98 (2007)
[Phys. At. Nucl. {\bf 70}, 93 (2007).\\
3. G. D. Alkhazov, T. Bauer, R. Bertini, {\it et al}.,  Nucl.
Phys. A {\bf 280}, 365 (1977).\\
4. G. D. Alkhazov, V. V. Sarantsev, Yad. Fiz. {\bf 75}, 1624
(2012) [Phys. At. Nucl. {\bf 75}, 1544 (2012)].\\
5. J. S. Al-Khalili and J. A. Tostevin, Phys. Rev. C {\bf 57}, 1846 (1998).\\
6. I. J. Thompson and M. V. Zhukov, Phys. Rev. C {\bf 49}, 1904 (1994).\\
7. G. D. Alkhazov, A. V. Dobrovolsky, P. Egelhof, {\it et al}.,
Nucl. Phys. A {\bf 712}, 269 (2002).\\
8. A. V. Dobrovolsky, G. D. Alkhazov, M. N. Andronenko, {\it et
al}., Nucl. Phys. A {\bf 766}, 1 (2006).\\

\end{document}